\documentclass[letterpaper, 10 pt, conference]{ieeeconf}  

\IEEEoverridecommandlockouts                              

\usepackage{epsfig} 
\usepackage{bm}
\usepackage{epsfig}
\usepackage{graphicx}
\usepackage{subfigure}
\usepackage{float}
\usepackage{cite}
\usepackage{url}
\usepackage{color}
\usepackage{balance}
\usepackage{mdwlist}
\usepackage{multirow}
\usepackage{threeparttable}
\let\labelindent\relax

\usepackage{enumitem}
\usepackage{amsmath}
\usepackage{amssymb}
\usepackage{stmaryrd}
\usepackage{booktabs}
\usepackage{siunitx}
\usepackage{amsmath}
\usepackage{threeparttable}
\usepackage{epsfig,amsmath,amsfonts}
\usepackage{amsfonts}

\usepackage{graphicx}
\usepackage{graphics}

\usepackage{caption}
\usepackage{subfigure}
\usepackage{stfloats}

\makeatletter
\newif\if@restonecol
\makeatother

\usepackage[ruled,vlined]{algorithm2e}
\usepackage{algpseudocode}
\usepackage{amsmath}

\newcommand{\thickhline}{\noalign{\hrule height 1.0pt}}

\newcommand{\parm}{{\xi}}
\newcommand{\vecpar}{\boldsymbol{\parm}}

\newcommand{\multiGPC}{\Psi }

\newcommand{\mat}[1]{\mathbf{#1}}

\RequirePackage{url}

\def\BibTeX{{\rm B\kern-.05em{\sc i\kern-.025em b}\kern-.08em
    T\kern-.1667em\lower.7ex\hbox{E}\kern-.125emX}}

\title{\LARGE \bf
Stochastic Model Predictive Control of Autonomous Systems with Non-Gaussian Correlated Uncertainty
}

\author{Huishan Chen and Zheng Zhang
\thanks{This work was supported by NSF CCF No. 1763699, NSF CAREER Award No. 1846476, and UCSB Academic Senate Faculty Research Grant.}
\thanks{Huishan Chen and Zheng Zhang are with Department of Electrical and Computer Engineering, University of California, Santa Barbara, CA 93106, USA.
        {\tt\small huishan\_chen@ucsb.edu, zhengzhang@ece.ucsb.edu.}}%
}

\begin{document}

\maketitle
\thispagestyle{empty}
\pagestyle{empty}

\begin{abstract}

Many systems such as autonomous vehicles and quadrotors are subject to parametric uncertainties and external disturbances. These uncertainties can lead to undesired performance degradation and safety issues. Therefore, it is important to design robust control strategies to safely regulate the dynamics of a system. This paper presents a novel framework for chance-constrained stochastic model predictive control of dynamic systems with non-Gaussian correlated probabilistic uncertainties. We develop a new stochastic Galerkin method to propagate the uncertainties using a new type of basis functions and an optimization-based quadrature rule. This formulation can easily handle non-Gaussian correlated uncertainties that are beyond the capability of generalized polynomial chaos expansions. The new stochastic Galerkin formulation enables us to convert a chance-constraint stochastic model predictive control problem into a deterministic one. We verify our approach by several stochastic control tasks, including obstacle avoidance, vehicle path following, and quadrotor reference tracking.

\end{abstract}

\section{INTRODUCTION}
\par
Model predictive control (MPC, also known as receding horizon control) has been widely used to predict and control the future events of a system in an uncertain environment. The representative applications of MPC include chemical process control~\cite{li2000robust}, economics~\cite{herzog2007stochastic}, path planning and obstacle avoidance of vehicles and robots~\cite{blackmore2011chance,guechi2018model,kim2013generalised,abbas2017obstacle,liu2015stochastic,huang2016model}, and air traffic management~\cite{kantas2009sequential}. One popular strategy for MPC is the robust MPC formulation. With the assumption that uncertainty belongs to a bounded set, robust MPC analyzes the stability and performance of the system against worst-case perturbations~\cite{bemporad1999robust,garatti2013modulating}. However, design based on worst-case uncertainties can be over-conservative in practice and may lead to in-feasibility in real applications. 
\par
In contrast to robust MPC, stochastic MPC with probabilistic constraints (also known as chance constraints) incorporates probabilistic descriptions of constraint violations, and allows for acceptable levels of risk during operations~\cite{blackmore2010probabilistic,primbs2009stochastic,schwarm1999chance}. Chance-constraint stochastic MPC avoids over-conservative decision making by directly incorporating the trade-offs between closed-loop performance and control constraints. One challenge in stochastic MPC is the computational cost of propagating uncertainties. A common approach is to use random sampling methods~\cite{vidyasagar2001randomized,shapiro2008stochastic} such as Monte Carlo. The system model is simulated repeatedly based on the samples to predict the time evolution of the uncertainty. Even though sampling-based approaches are applicable to most problems, these techniques can be extremely expensive due to the large number of samples required to achieve accurate uncertainty propagation. 

\par

Polynomial chaos expansions provide an efficient alternative to propagate the uncertainties through the system dynamics~\cite{fisher2011optimal,fisher2008stability,mesbah2014stochastic,boutselis2019numerical}.  Based on the generalized polynomial chaos theory~\cite{xiu2002wiener}, some popular distributions such as Gaussian distributions and independent uniform distributions have been handled successfully in stochastic MPC. These techniques have been applied in the trajectory optimization of quadrotors and oscillators~\cite{boutselis2019numerical}, and chemical process control~\cite{mesbah2014stochastic}. However, uncertainties are often non-Gaussian correlated in practice and do not follow the common probability distributions described in~\cite{xiu2002wiener}. Probabilistic reachable sets have been used by~\cite{hewing2018recursively} to overestimate the impact caused by correlated external disturbances. However, accurate uncertainty propagation for systems with non-Gaussian correlated parametric uncertainties has not been considered.

\par
\textbf{Paper Contributions.} 
Inspired by a new stochastic collocation technique described in~\cite{cui2018stochastic,cui2018stochasticarticle}, this paper presents an efficient stochastic MPC framework that can deal with more realistic non-Gaussian correlated uncertainties. Our specific contributions include:
\begin{itemize}
    \item A novel stochastic Galerkin method to propagate the uncertainties of a dynamic system with non-Gaussian correlated uncertainties. Traditional stochastic Galerkin methods~\cite{ghanem2003stochastic} assume that uncertain parameters are mutually independent or Gaussian correlated, thus cannot handle the challenging cases presented in this paper.
    \item We apply our stochastic Galerkin formulation to develop a chance-constraint stochastic model predictive control solver that can handle non-Gaussian correlated uncertainties efficiently and accurately.  
    \item Numerical results. We verify our proposed framework by solving obstacle avoidance, vehicle path following, and quadrotor reference tracking problems that involve non-Gaussian correlated uncertainties. 
\end{itemize}

\section{CHANCE CONSTRAINED STOCHASTIC MPC}
\subsection{Problem Formulation}
 Consider a stochastic discrete-time linear system:
    \begin{equation}
      \label{eq:system}   \mathbf{x}_{t+1} = \mathbf{A}(\vecpar)\mathbf{x}_t + \mathbf{B}(\vecpar)\mathbf{u}_t + \mathbf{D}(\vecpar) \boldsymbol{\omega}_t, \forall t=0,\ldots,T-1,
    \end{equation}
where $\mathbf{x}_t \in \mathbb{R}^{n_{\mathbf{x}}}$, $\mathbf{u}_t \in \mathbb{R}^{n_{\mathbf{u}}}$,  and $\boldsymbol{\omega}_t \in \mathbb{R}^{n_{\boldsymbol{\omega}}}$ denote the stochastic system states, inputs, and disturbances at current time, respectively. The system has probabilistic uncertainty in the system parameters, characterized by $\mathbf{A}(\vecpar)$, $\mathbf{B}(\vecpar)$, and $\mathbf{D}(\vecpar)$, which depend on the uncertain parameter $\vecpar \in \mathbb{R}^d$. Assume that the joint probability density function (PDF) of random variable $\vecpar$ is known as ${\rm PDF}(\vecpar)$. Assume that $\boldsymbol{\omega}_t$ depends on $\vecpar$ and its PDF is also known at every time point $t$. Because of the stochasiticity of parameters and disturbances, the system trajectory will also be stochastic.

\par
Chance constraint is an efficient technique for solving control or optimization problems with uncertainty~\cite{li2008chance}. Unlike a worst-case optimization where all constraints are satisfied with probability $1$, a chance-constraint optimization ensures the probability of satisfying a control/optimization constraint is above a certain confidence level $\beta$:
\begin{equation}
   \text{Pr}\left[ \mathbf{x}_t  \notin \mathcal{F}_{\mathbf{x}} \right] \geq \beta,
\end{equation}
where $\mathcal{F}_{\mathnormal{x}}$ denotes the forbidden (e.g., unsafe) region for the state vector $\mathbf{x}_t$, and $\beta$ is the confidence level.

\par
In finite-horizon stochastic MPC, the goal is to determine a control policy $\boldsymbol{\mu}_T \triangleq (\mathbf{u}_0,..., \mathbf{u}_{T-1})$ that can drive the state vector $\mathbf{x}_t$ to have a desirable statistical performance. The optimal control can be solved from the following optimization:
\begin{equation}
\label{eq:MPC_1}
\begin{aligned}
       \text{{\bf Problem  1: }} & \text{Stochastic  MPC  with  chance  constraints}\\
        &\underset{\boldsymbol \mu_T}{\text{min}} \ \mathbb{E}\left[  \sum_{t=1}^T \mathbf{x}_t^T\mathbf{Q}_t \mathbf{x}_t + \mathbf{u}_{t-1}^T\mathbf{R}_t \mathbf{u}_{t-1} \right]\\
        & \text{s.t.} \ \mathbf{x}_{t+1} = \mathbf{A}(\vecpar)\mathbf{x}_t + \mathbf{B}(\vecpar)\mathbf{u}_t + \mathbf{D}(\vecpar)\boldsymbol{\omega}_t,\\
        &  \ \ \ \ \text{Pr}\left[ \mathbf{x}_t  \notin \mathcal{F}_{\mathbf{x}} \right] \geq \beta,\\
        &  \ \ \ \ \mathbf{u}_t \in \mathcal{U}, \;  \mathbf{x}_0  = \mathbf{x}_{\rm init}\left( \vecpar \right),\\
        &  \ \ \ \ \vecpar \sim \mathnormal{f}_{\vecpar} , \ \ \ \ \boldsymbol{\omega}_t \sim \mathnormal{f}_{\boldsymbol{\omega}_t},  
\end{aligned}
\end{equation}
where the initial condition $\mathbf{x}_{\rm init}\left(   \vecpar \right)$ is given and can be uncertain, $\mathcal{U} \subset \mathbb{R}^{n_{\mathbf{u}}}$ is the compact set of input constraints, $\mathbf{Q}_t$ and $\mathbf{R}_t$ are positive definite weight matrices, $\mathnormal{f}_{\vecpar}$ and $\mathnormal{f}_{\boldsymbol{\omega}_t}$ denote the PDFs of $\vecpar$ and $\boldsymbol{\omega_t}$, respectively. 
 
\subsection{Polynomial Chaos-Based Stochastic MPC}
In order to obtain  $\mathbf{x}_t(\vecpar)$ in the MPC, one can employ a truncated generalized polynomial-chaos expansion~\cite{xiu2002wiener}:
\begin{equation}\label{eq:gPC_expansion}
  \mathbf{x}_t(\vecpar) \approx \sum\limits_{k=1}^{N_p} \mathbf{c}_{k,t}\multiGPC_k(\vecpar),
\end{equation}
where $\multiGPC_{k}(\vecpar)$ is the orthonormal multivariate basis function satisfying $\left\langle \multiGPC_k(\vecpar), \multiGPC_j(\vecpar) \right\rangle=\mathbb{E}\left[ \multiGPC_k(\vecpar)\multiGPC_j(\vecpar)  \right] = \delta_{k,j}$. $\delta_{k,j} = 1$ if $k=j$ and otherwise $\delta_{k,j} = 0$. If we upper bound the total polynomial order by $p$, then the total number of basis functions is $N_p = \frac{(p+d)!}{(p!d!)}$. With the above truncated generalized polynomial-chaos expansion, the time-dependent deterministic coefficients $\mathbf{c}_{k,t}$ can be computed efficiently via numerical schemes such as stochastic collocation~\cite{xiu2005high}, stochastic Galerkin~\cite{ghanem2003stochastic}, or stochastic testing~\cite{zhang2013stochastic}. Finally, the formulation in~\eqref{eq:MPC_1} can be converted to a deterministic one and solved efficiently~\cite{mesbah2014stochastic,kim2013generalised,fisher2011optimal}. 

{\bf Limitations.} Despite their high efficiency, existing polynomial chaos-based stochastic MPC~\cite{fisher2011optimal,fisher2008stability,mesbah2014stochastic,boutselis2019numerical} is limited by a strong assumption of generalized polynomial chaos~\cite{xiu2002wiener}: the uncertain parameters $\vecpar$ should be mutually independent or just Gaussian correlated (which can be easily de-correlated). In practice, $\vecpar$ can be non-Gaussian correlated, thus the basis functions proposed in~\cite{xiu2002wiener} cannot be employed. Furthermore, the numerical solvers such as stochastic collocation~\cite{xiu2005high} and stochastic Galerkin~\cite{ghanem2003stochastic} reply on fast numerical integration rules such as Gaussian quadrature~\cite{golub1969calculation} or spare grid~\cite{nobile2008sparse}, which fail for non-Gaussian correlated parameters as well.

\section{Stochastic Galerkin with Non-Gaussian Correlated Uncertainties}

In order to handle non-Gaussian correlated uncertainties in a stochastic MPC problem efficiently, this section proposes a new stochastic Galerkin formulation. Our formulation extends the work of~\cite{cui2018stochastic,cui2018stochasticarticle}. Note that the methods in~\cite{cui2018stochastic,cui2018stochasticarticle} are stochastic collocation, and they are less efficient than our method when solving time-evolving problems. 

\par
\subsection{Deterministic Formulation via Stochastic Galerkin}

\par
Since the system parameters and disturbances are stochastic variables, we express them with truncated expansion of stochastic basis functions:
    \begin{equation}
        \mathbf{u}_t(\vecpar) \approx  \sum\limits_{k = 1}^{N_p} \mathbf{h}_{k,t}\multiGPC_{k}(\vecpar), \; \; \boldsymbol{\omega}_t\approx  \sum\limits_{k = 1}^{N_p} \mathbf{w}_{k,t}\multiGPC_{k}(\vecpar).
    \end{equation}
If the control input $\mathbf{u}_t$ is deterministic, then $\mathbf{h}_{1,t} = \mathbf{u}_t$ and all other coefficients $\mathbf{h}_{k,t}(k \ne 1)$ are set to zero. Based on the above expansions, \eqref{eq:system} can be rewritten as:
\begin{equation}
\label{eq:newSystem}
\begin{aligned}
    &\sum\limits_{k = 1}^{N_p} \mathbf{c}_{k,t+1}\multiGPC_{k}(\vecpar)  \approx \sum\limits_{k= 1}^{N_p} \mathbf{c}_{k,t}\mathbf{A}(\vecpar)\multiGPC_{k}(\vecpar) \\
    & + \sum\limits_{k=1}^{N_p} \mathbf{h}_{k,t}\mathbf{B}(\vecpar)\multiGPC_{k}(\vecpar)+\sum\limits_{k =1}^{N_p} \mathbf{w}_{k,t}\mathbf{D}(\vecpar)\multiGPC_{k}(\vecpar).\\
 \end{aligned}
\end{equation}

Stochastic Galerkin can reformulate the stochastic system into a deterministic one by enforcing the residual of \eqref{eq:newSystem} orthogonal to each stochastic basis function. By Galerkin projection, for each $j = 1,...,N_p$, we have
\begin{align}
\label{eq:projection}
 \mathbf{c}_{j,t+1} & \approx  \sum\limits_{k= 1}^{N_p} \mathbf{c}_{k,t} \Bigg\langle \mathbf{A}(\vecpar) \multiGPC_{k}(\vecpar), \multiGPC_{j}(\vecpar) \Bigg\rangle \nonumber \\
    &+   \sum\limits_{k=1}^{N_p} \mathbf{h}_{k,t} \Bigg\langle \mathbf{B}(\vecpar)\multiGPC_{k}(\vecpar), \multiGPC_{j}(\vecpar) \Bigg\rangle \nonumber \\
    & + \sum\limits_{k =1}^{N_p} \mathbf{w}_{k,t} \Bigg\langle \mathbf{D}(\vecpar)\multiGPC_{k}(\vecpar), \multiGPC_{j}(\vecpar) \Bigg\rangle , 
\end{align}
where the inner product of  $f_1(\vecpar)$ and $f_2(\vecpar)$ equals:
\begin{equation}
    \left \langle f_1(\vecpar), f_2 (\vecpar )\right \rangle= \int \limits_{\mathbb{R}^d} f_1(\vecpar) f_2 (\vecpar ) {\rm PDF} (\vecpar) d\vecpar.   
\end{equation}
\par

Let us define
\begin{equation}
\begin{aligned}
    \hat{\mat{x}}_t & =\begin{bmatrix}
    \mat{c}_{1,t} & \cdots & \mat{c}_{N_p,t}
    \end{bmatrix}^T, \\
    \hat{\mat{u}}_t & =\begin{bmatrix}
    \mat{h}_{1,t} & \cdots & \mat{h}_{N_p,t}
    \end{bmatrix}^T,\\
        \hat{\boldsymbol{\omega}}_t &=\begin{bmatrix}
    \mat{w}_{1,t} & \cdots & \mat{w}_{N_p,t}
    \end{bmatrix}^T.
\end{aligned}
\end{equation}
The Galerkin projection in Equation~\eqref{eq:projection} generates the following deterministic dynamic system:
    \begin{equation}
    \label{eq:deterministic}
      \hat{\mathbf{x}}_{t+1} = \hat{\mathbf{A}}\hat{\mathbf{x}}_t + \hat{\mathbf{B}}\hat{\mathbf{u}}_t +  \hat{\mathbf{D}}\hat{\boldsymbol{\omega}}_t,
    \end{equation}
where the vectors $\hat{\mathbf{x}}_t \in \mathbb{R}^{n_{\mathbf{x}}N_p}$ and $\hat{\boldsymbol{\omega}}_t \in \mathbb{R}^{n_{\boldsymbol{\omega}}N_p}$ contain their truncated expansion coefficients at time $t$. The matrices $\hat{\mathbf{A}} \in \mathbb{R}^{n_{\mathbf{x}}N_p \times n_{\mathbf{x}}N_p}$, $\hat{\mathbf{B}} \in \mathbb{R}^{n_\mathbf{x}N_p \times n_{\mathbf{u}}N_p}$, $\hat{\mathbf{D}} \in \mathbb{R}^{n_\mathbf{x}N_p \times n_{\boldsymbol{\omega}}N_p}$ are matrices with $N_p \times N_p$ blocks. Their $(k,j)$-th blocks can be computed off-line by the following formula: 
\begin{equation}
\label{eq:matBlock}
\begin{aligned}
 \hat{\mat{A}}_{k,j} & =  \Bigg\langle \mathbf{A}(\vecpar) \multiGPC_{k}(\vecpar), \multiGPC_{j}(\vecpar) \Bigg\rangle \in \mathbb{R}^{n_{\mat{x}} \times n_{\mat{x}} },  \\
   \hat{\mat{B}}_{k,j} &=    \Bigg\langle \mathbf{B}(\vecpar)\multiGPC_{k}(\vecpar), \multiGPC_{j}(\vecpar) \Bigg\rangle  \in \mathbb{R}^{n_{\mat{x}} \times n_{\mat{u}} } ,  \\
   \hat{\mat{D}}_{k,j} & = \Bigg\langle \mathbf{D}(\vecpar)\multiGPC_{k}(\vecpar), \multiGPC_{j}(\vecpar) \Bigg\rangle \in \mathbb{R}^{n_{\mat{x}} \times n_{\boldsymbol{\omega} } }. 
\end{aligned}
\end{equation}

As shown in Eq.~\eqref{eq:matBlock}, we need to evaluate a set of inner products in order to set up the deterministic dynamic system model~\eqref{eq:deterministic}. The inner product calculation requires an accurate numerical integration method. For a general smooth function $g(\vecpar)$, its integration can be evaluated with a set of quadrature points and weights $\{ \vecpar_l, w_l \}_{l=1}^M$ as follows:
\begin{equation}
\label{eq:quadrature}
    \int \limits_{\mathbb{R}^d} g(\vecpar) {\rm PDF}(\vecpar)d \vecpar \approx \sum \limits_{l=1}^M g(\vecpar_l) w_l.
\end{equation}

There exist two challenges in the stochastic Galerkin formulation. Firstly, how shall we choose the stochastic basis functions $\{ \multiGPC_{k}(\vecpar)\}$ such that they can capture the impacts caused by non-Gaussian correlated uncertainties? Secondly, how shall we choose the quadrature points and weights in the correlated parameter space? We will employ the methods of~\cite{cui2018stochastic,cui2018stochasticarticle} in our stochastic Galerkin formulation, as elaborated below.

\par

\subsection{Selection of Basis Functions}
\label{subsec:basis}

The generalized polynomial chaos~\cite{xiu2002wiener} cannot be used when the parameters $\vecpar$ are non-Gaussian correlated. In this case, orthonormal basis functions $\{\Psi _k\left( \vecpar \right)\}$ can be obtained based on a multi-variate Gram-Schmidt process~\cite{cui2018stochastic,cui2018stochasticarticle}:
\begin{enumerate}
    \item Reorder the monomials $p_{\bm{\alpha}}(\vecpar) = \xi_1^{\alpha_1}...\ \xi_d^{\alpha_d}$ in the graded lexicographic order and denote them as $\{ p_k (\vecpar)\}$.
    \item Set $\Psi_1(\vecpar) = 1 $, calculate the orthonormal polynomials $\left\{  \Psi_k(\vecpar) \right\}_{k=1}^{N_p}$ recursively:
        \begin{equation}
        \label{eq:basisConstruct}
          \begin{aligned}  &\hat{\Psi}_k(\vecpar) = p_k(\vecpar) - \sum_{i = 1}^{k-1} \mathbb{E}\left[ p_k(\vecpar)\Psi_i(\vecpar)\right]\Psi_i(\vecpar),\\
           &\Psi_k(\vecpar) = \frac{\hat{\Psi}_k(\vecpar)}{\sqrt{\mathbb{E}\left[ \hat{\Psi}_k^2(\vecpar)   \right]}}, \ \ \ k = 2,...,N_p.\\
          \end{aligned}
        \end{equation}
\end{enumerate}

\par
Once the dynamic system \eqref{eq:deterministic} is set up and solved, the statistics of the stochastic variable $\mathbf{x}_t(\vecpar)$ can be efficiently computed. For example, the mean and variance of $\mathbf{x}_t(\vecpar)$ are 
    \begin{equation}
    \label{eq:meanVar}
        \mathbb{E}\left[ \mathbf{x}_t(\vecpar)   \right] =\mathbf{c}_{1,t}, \; \text{Var}\left[ \mathbf{ x}_t(\vecpar)   \right] = \sum\limits_{k = 2 }^{N_p} \mathbf{c}_{k,t}^2 .
    \end{equation}

\subsection{Optimization Based Quadrature Rule}
To evaluate the matrices in \eqref{eq:matBlock}, we need some quadrature points and samples such that the integration in \eqref{eq:quadrature} can be calculated with a high accuracy for a general smooth function $g(\vecpar)$. We employ the optimization-based quadrature~\cite{cui2018stochastic,cui2018stochasticarticle} for non-Guassian correlated parameters $\vecpar$.

Assume that we want to get an (almost) exact integration for a polynomial function $g(\vecpar)$ bounded by degree $2p$. In this case, we can express $g(\vecpar)$ with $N_{2p} = \frac{(2p+d)!}{(2p!d!)}$ basis functions bounded by order-$2p$. The quadrature nodes and weights $\left\{ \vecpar_l, \mathnormal{w}_l\right\}_{l=1}^M$ are determined by enforcing exact integration over these basis functions:
\begin{equation}\label{eq:integration}
    \mathbb{E}\left[  \Psi_k(\vecpar) \right] = \sum_{l=1}^M \Psi_k(\vecpar_l)\mathnormal{w}_l, \ \ \ \forall k = 1,...,N_{2p}.
\end{equation}
Based on the orthonormal condition for $\Psi_k(\vecpar)$, it is easy to show that $\mathbb{E}\left[ \Psi_k(\vecpar) \right] = \mathbb{E}\left[ \Psi_k(\vecpar)\Psi_1(\vecpar) \right] = \delta_{1k}$, where $\delta_{1k} = 1$ if $k=1$ and otherwise $\delta_{1k} = 0$. As a result, \eqref{eq:integration} is reformulated into a nonlinear least square optimization:
\begin{equation}\label{eq:nonlinearopt}
    \underset{\bar{\vecpar},\mathbf{w}}{\text{min}} \ \left\| \mathbf{\Phi}\left( \bar{\vecpar}  \right)\mathbf{w} - \mathbf{e}_1 \right\|_2^2,
\end{equation}
where $\bar{\vecpar} = \left[ \vecpar_1,...,\vecpar_M  \right]^T \in \mathbb{R}^{M \times d}$, $\mat{w} = \left[ w_1,...,w_M  \right]^T \in \mathbb{R}^{M}$, 
$\mat{e}_1= \left[ 1,0,...,0  \right] \in \mathbb{R}^{N_{2p}}$, $\left[\boldsymbol{\Phi}(\bar{\vecpar})\right]_{kl} = \Psi_k(\vecpar_l)$  and $\boldsymbol{\Phi}(\bar{\vecpar}) \in \mathbb{R}^{N_{2p} \times M}$. A block coordinate-descent method was used in~\cite{cui2018stochastic,cui2018stochasticarticle} to solve this optimization problem. 

 In order to make the algorithm more efficient, a weighted clustering approach is used in order to automatically reduce the number of quadrature points and achieve desired level of accuracy. Specifically, once \eqref{eq:nonlinearopt} is solved, $M$ is reduced by one by a weighted clustering method~\cite{cui2018stochasticarticle}, then \eqref{eq:nonlinearopt} is solved again using the reduced samples and weights. This process is repeated until a further reduction leads to failures when solving~\eqref{eq:nonlinearopt}~\cite{cui2018stochastic,cui2018stochasticarticle}.

\section{APPLICATION IN STOCHASTIC MPC}

In this section, we describe how to perform stochastic MPC using the deterministic system \eqref{eq:deterministic} obtained by the above stochastic Galerkin formulation.

\subsection{Minimum Expectation Control}
We consider the minimum expectation control problem:
    \begin{equation}
    \label{eq:cost}
        \underset{\boldsymbol \mu_T}{\text{min}} \ \mathbb{E}\left[  \sum_{t=1}^T \mathbf{x}_t^T\mathbf{Q}_t \mathbf{x}_t + \mathbf{u}_{t-1}^T \mathbf{R}_t \mathbf{u}_{t-1} \right],
    \end{equation}
where $\mathbf{Q}_t$ and $\mathbf{R}_t$ are user defined state and input weight matrices satisfying $\mathbf{Q}_t = \mathbf{Q}^T_t > 0$ and $\mathbf{R}_t = \mathbf{R}^T_t > 0$. With a truncated expansion using the stochastic basis functions described in Section~\ref{subsec:basis}, the term $\mathbb{E}\left[\mathbf{x}^T_t  \mathbf{x}_t  \right]$ is obtained as 
    \begin{equation}
    \label{eq:exp_square}
        \mathbb{E}\left[\mathbf{x}^T_t  \mathbf{x}_t \right] = \hat{\mathbf{x}}_t^T\left( \mathbf{I}_{n_{\mathbf{x}}} \otimes \mat{V}  \right) \hat{\mathbf{x}}_t,
    \end{equation}
where $\mathbf{I}_{n_{\mathbf{x}}} \in \mathbb{R}^{n_{\mathbf{x}} \times n_{\mathbf{x}}}$ is an identity matrix,  $\otimes$ is a Kronecker product, and $\mat{V} \in \mathbb{R}^{N_p \times N_p} = \left\{ v_{ij} \right\}$ with $v_{ij} = \sum_{l=1}^M \Psi_i(\vecpar_l)\Psi_j(\vecpar_l)\mathnormal{w}_l$. Based on \eqref{eq:exp_square} the cost function in \eqref{eq:cost} can now be written in terms of deterministic variables:
    \begin{equation}
      \label{eq:deterministicObjective}  \underset{\boldsymbol \mu_T}{\text{min}} \ \  \sum_{t=1}^T \hat{\mathbf{x}}_t\hat{\mathbf{Q}}_t\hat{\mathbf{x}}_t+ \hat{\mathbf{u}}_{t-1}\hat{\mathbf{R}}_t\hat{\mathbf{u}}_{t-1},
    \end{equation}
where $\hat{\mathbf{Q}}_t = \mathbf{Q}_t \otimes \mathbf{V}$ and  $\hat{\mathbf{R}}_t = \mathbf{R}_t \otimes \mathbf{V}$.

\subsection{Deterministic Optimization for Stochastic MPC}

\begin{algorithm}[t]
\caption{Chance-Constrained Stochastic MPC solver with Non-Gaussian Correlated Uncertainties}
\label{alg:the_alg}
\SetKwInput{KwInput}{Input} 
\SetKwInput{KwOutput}{Output} 

\KwInput{PDF of the non-Gaussian correlated random parameters $\vecpar$ and $\bm{\omega}_t$, system parameters $\mathbf{A}(\vecpar)$, $\mathbf{B}(\vecpar)$, and $\mathbf{D}(\vecpar)$, polynomial order $p$, final time $T$, initial condition $\mathbf{x}_{\rm init}(\vecpar)$, weight matrices $\mathbf{Q}_t$ and $\mathbf{R}_t$, input constraints set $\mathcal{U}$, forbidden regions $\mathcal{F}_{\mathbf{x}}$, confidence level $\beta$. }
\BlankLine
\begin{enumerate}
    \item[1.]  Construct the basis functions $\Psi _k\left( \vecpar \right)$ by \eqref{eq:basisConstruct}.
    \item[2.] Construct the quadrature points $\vecpar_l$ and weights $\mathnormal{w}_l$ by \eqref{eq:nonlinearopt} and weighted clustering.
    \item[3.] Set up the deterministic system \eqref{eq:deterministic} via \eqref{eq:matBlock}.
    \item[4.] Convert the objective function and chance constraints into deterministic ones by \eqref{eq:deterministicObjective} and \eqref{eq:deterministicConstraint}, respectively.
    \item[5.] Solve the optimization problem \eqref{eq:deterministicSurrogate}. 
\end{enumerate}

\KwOutput{Optimized input signals $\mathbf{u}^{*}_t$, and the time-dependent stochastic state $\mathbf{x}^{*}_t$.
}
\end{algorithm}

Assume that the forbidden region in~\eqref{eq:MPC_1} is described by some inequalities, and the chance constraints can be written in the following form:
    \begin{equation}
        \text{Pr}\left[ g(\mathbf{x}_t(\vecpar)) \leq 0  \right] \geq \beta.
    \end{equation}
This probabilistic constraint can be reformulated as some deterministic constraints ~\cite{calafiore2005distributionally} by using the mean and variance of $g(\mathbf{x}_t(\vecpar))$:
    \begin{equation}
    \label{eq:deterministicConstraint}
        \mathbb{E}\left[ g(\mathbf{x}_t(\vecpar)) \right] + \kappa_{1-\beta}\sqrt{\text{Var}\left[ g(\mathbf{x}_t(\vecpar)) \right]} \geq 0,
    \end{equation}
where $\mathbb{E}\left[\cdot \right]$ and $\text{Var}\left[\cdot \right]$ are obtained via \eqref{eq:meanVar}. The constant $\kappa_{1-\beta}$ is chosen as $\kappa_{1-\beta} = \sqrt{\beta/\left(1-\beta\right)}$.

With \eqref{eq:deterministic}, \eqref{eq:deterministicObjective} and \eqref{eq:deterministicConstraint}, we are ready to re-write Problem 1 as a deterministic optimization problem:

{\bf Problem  2}: Deterministic Optimization for Stochastic MPC:
    \begin{equation}
    \label{eq:deterministicSurrogate}
    \centering
   \begin{aligned}
        & \underset{\boldsymbol \mu_T}{\text{min}} \ \  \sum_{t=1}^T \hat{\mathbf{x}}_t\hat{\mathbf{Q}}_t\hat{\mathbf{x}}_t+ \hat{\mathbf{u}}_{t-1}\hat{\mathbf{R}}_t\hat{\mathbf{u}}_{t-1}\\
        & \text{s.t.} \ \  \hat{\mathbf{x}}_{t+1} = \hat{\mathbf{A}}\hat{\mathbf{x}}_t + \hat{\mathbf{B}}\hat{\mathbf{u}}_t +  \hat{\mathbf{D}}\hat{\boldsymbol{\omega}}_t,\\
        &  \ \ \ \ \ \mathbb{E}\left[ g(\mathbf{x}_t(\vecpar)) \right] + \kappa_{1-\beta}\sqrt{\text{Var}\left[ g(\mathbf{x}_t(\vecpar)) \right]} \geq 0,\\
      &  \ \ \ \ \  \mathbf{u} \in \mathcal{U},\;\;  \hat{\mathbf{x}}_0 = \hat{\mathbf{x}}_{\rm init},\\
     \end{aligned}
    \end{equation}
where $\hat{\mathbf{x}}_{\rm init}$ is the expanded initial condition containing the expansion coefficients.

The whole framework is summarized in Alg.~\ref{alg:the_alg}.


\section{NUMERICAL RESULTS}
In order to verify our proposed chance constrained stochastic MPC method, we implement the algorithm
in MATLAB on a Windows Desktop Workstation with 8-GB
RAM and a 3.4-GHz CPU.

\subsection{Obstacles Avoidance}
\par
\begin{figure}[t]
    \centering
    \includegraphics[width=3.2in]{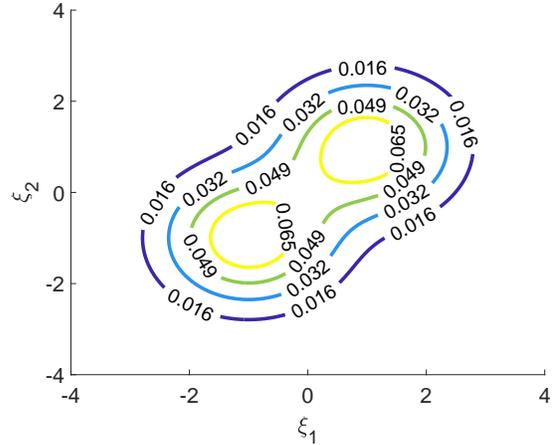}
    \caption{Joint PDF of random variables $\xi_1$ and $\xi_2$ for the obstable avoidance example.}
    \label{fig:jointPDF}
\end{figure}
\begin{figure*}[t]
\centering
  \subfigure{
        \includegraphics[width=3in]{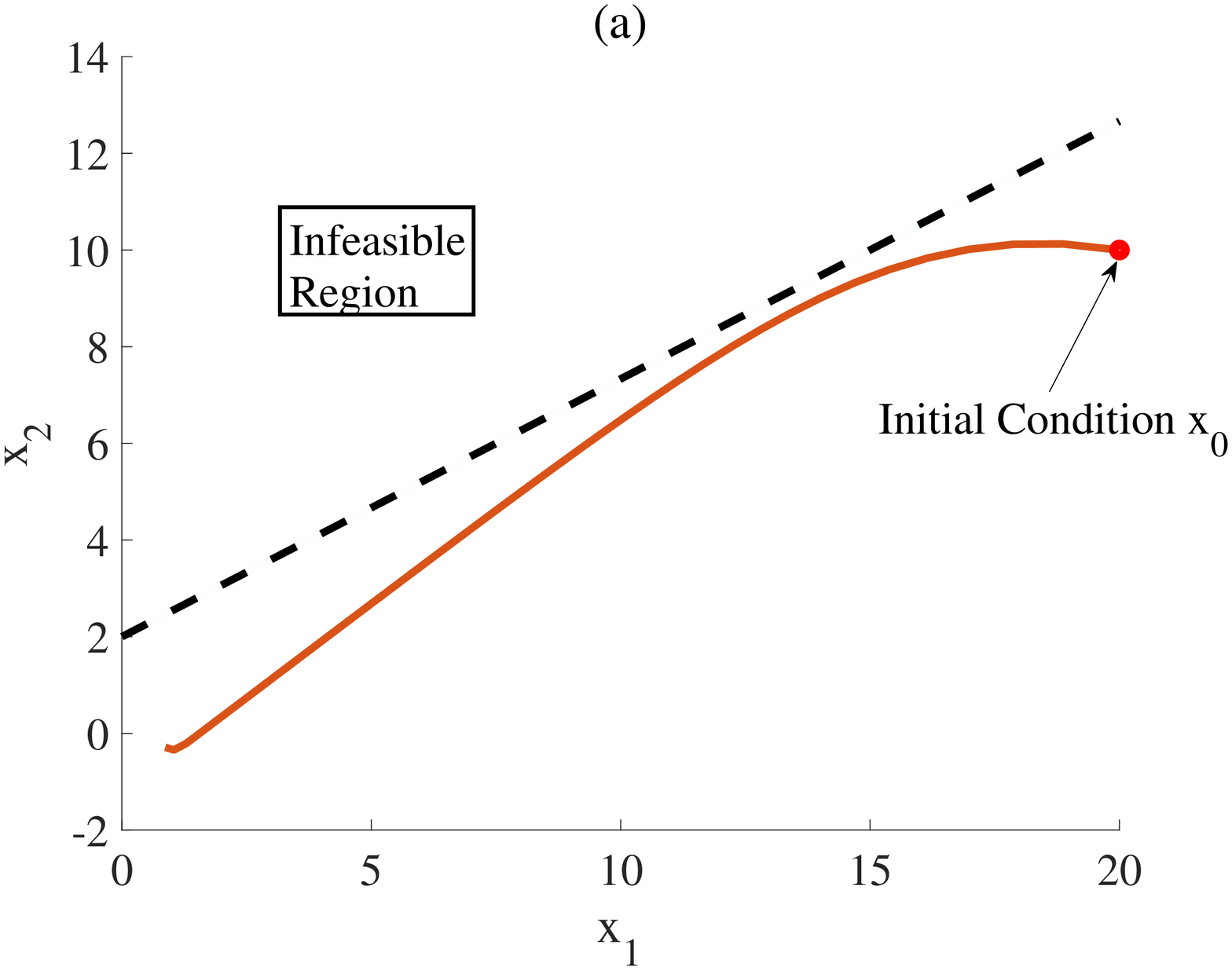}
        
        \label{fig:NoUncertaintyObstacle}
        }
\hspace{0.5pt}
  \subfigure{
        \includegraphics[width=3in]{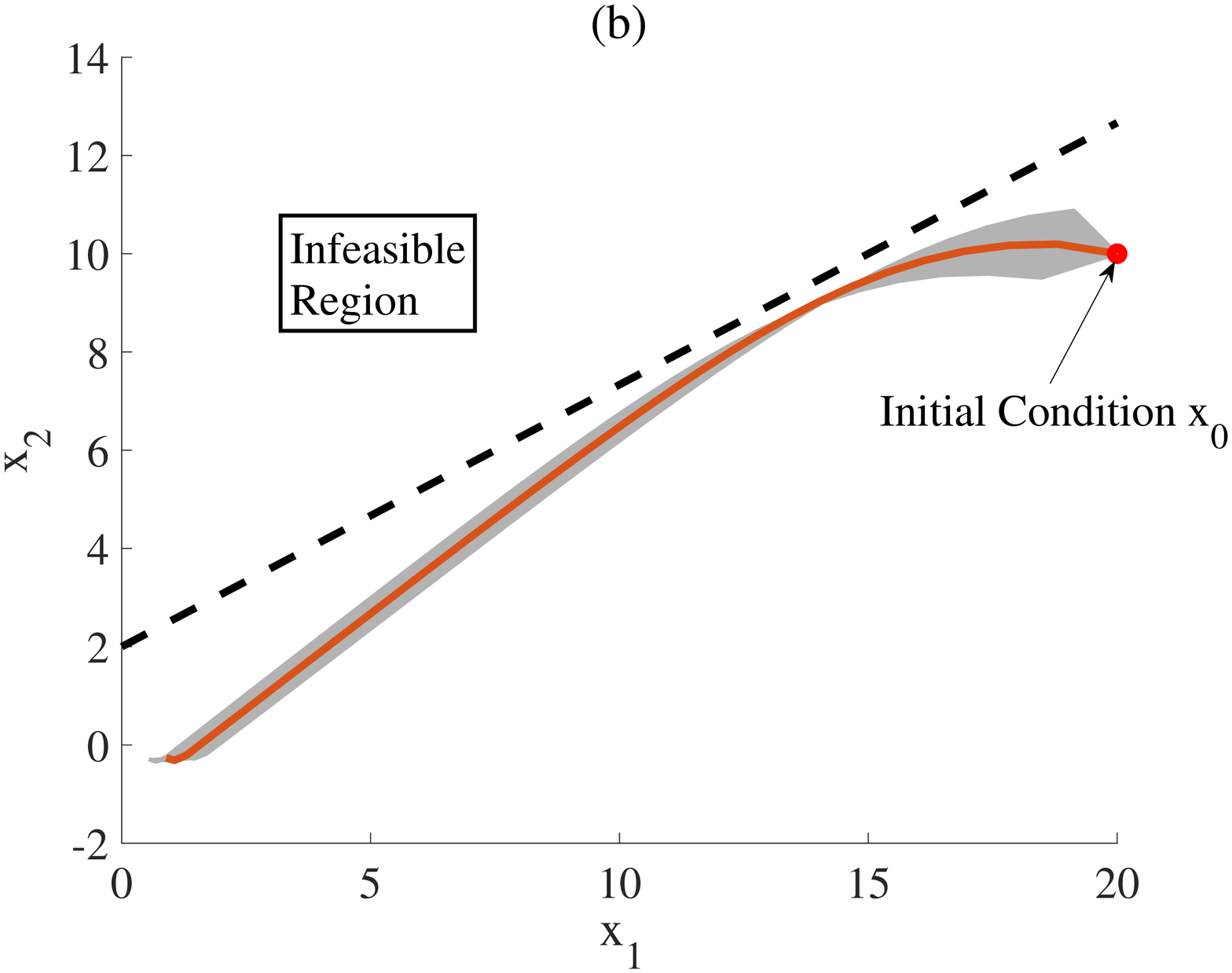}
       
        \label{fig:ObstacleWithUncertainty}
        }
    \caption{Controlled trajectories for the obstacles avoidance experiment. (a): without parameter uncertainty; (b): with non-Gaussian correlated uncertainty. Red line shows the controlled trajectory or its mean value, black dotted line separates the feasible and infeasible regions, and grey lines show the trajectory uncertainties.} 
    \label{fig:obstacleavoidance}
\end{figure*}

Consider a stochastic linear time-invariant  system
    \begin{equation}
        \mathbf{x}_{t+1} = \begin{bmatrix}
0.9+\rho_1\xi_1 & 0.1\\ 
0.1 & 0.85
\end{bmatrix}\mathbf{x}_{t} + \begin{bmatrix}
0.25-\rho_1\xi_1 \\ 
0.75+ \rho_2\xi_2
\end{bmatrix}\mathbf{u}_{t},
    \end{equation}
with initial condition $\mathbf{x}_0 = \left[ 20,10\right]^T$. The non-Gaussian correlated random variables $\xi_1$ and $\xi_2$ obey a Gaussian-mixture distribution, as shown in Fig. \ref{fig:jointPDF}. Here $\rho_1$ and $\rho_2$ are the weights on the random variables. The prediction horizon is $T = 4$ sec, and input constraint $\mathbf{u}_t \in \left[-0.5,0.5\right]$. The constraints describe an infeasible region shown in Fig. \ref{fig:NoUncertaintyObstacle}. 

\par
Consider $\mathbf{Q}_t = \text{diag}\left( 100, 100\right)$ and $\mathbf{R}_t = 1$ for all $t$, we use Alg.~\ref{alg:the_alg} to solve the problem with a $99\%$ level of confidence for avoiding the infeasible region. The resultant controlled trajectory generated without parameter uncertainty, i.e., $\rho_1 = \rho_2 = 0$, is shown in Fig. \ref{fig:NoUncertaintyObstacle}, which stays in the feasible region as desired. Consider the system with parameter uncertainty, i.e., $\rho_1 = 0.001$ and $\rho_2 = 0.05$. Fig. \ref{fig:ObstacleWithUncertainty} shows the controlled system trajectories generated with our proposed framework, where the red line shows the mean trajectory obtained via \eqref{eq:meanVar}. 



\begin{figure*}[tp]
\begin{equation}
    \label{eq: vechicle}
    \begin{aligned}
        &\dot{\mathbf{x}}(t) = \mathbf{A}\mathbf{x}(t) + \mathbf{B}\mathbf{u}(t), \; \;\;  \mathbf{y}(t) = \mathbf{C}\mathbf{x}(t),\\
        & \mathbf{A} = \begin{bmatrix}
        0 &1 &0 &0\\
        0 & -\frac{2C_f+2C_r}{mV_x} & \frac{2C_f+2C_r}{m} & -\frac{2aC_f+2bC_r}{mV_x} \\
        0 &0 &0 &1\\
        0 & -\frac{2aC_f+2bC_r}{I_{zz}V_x} & \frac{2aC_f-2bC_r}{I_{zz}} & -\frac{2a^2C_f+2b^2C_r}{I_{zz}V_x}\end{bmatrix},\;\;
        \mathbf{B} = \begin{bmatrix}
        0 & 0 \\ \frac{2C_f}{m} & \frac{2aC_f-2bC_r}{mV_x}-V_x \\ 0 & 0 \\ \frac{2aC_f}{I_{zz}} & \frac{2a^2C_f+2b^2C_r}{I_{zz}V_x}
        \end{bmatrix},\\
        & \mathbf{C} = \begin{bmatrix} 1 & 0& 0& 0 \end{bmatrix},\;\; \mathbf{x}(t) = \begin{bmatrix}  
        e_1(t) & \dot{e}_1(t) & e_2(t) & \dot{e}_2(t)
        \end{bmatrix}^T, \;\;  \mathbf{u}(t) = \begin{bmatrix} \delta(t) & \dot{\psi}_{\rm des}(t) \end{bmatrix}.
        \end{aligned}
    \end{equation}  
    \hrule
\end{figure*}

\subsection{Vehicle Path-following Problem}

Path-following control aims to enforce the system states converge to a given path, without temporal specifications~\cite{lapierre2006nonsingular,aguiar2007trajectory,walters2013online,lapierre2008robust}. We consider the vehicle path-following model in~\cite{lee2018analysis}, where the state variables include position error, orientation error, and their first order derivative with respect to time. The dynamic model for vehicle path following is generally nonlinear~\cite{rajamani2011vehicle,lee2018analysis}. With the assumption that the vehicle is moving at a constant speed of 20 m/s and the influence of road bank angle is neglected~\cite{lee2018analysis}, the vehicle can be described by the linear state-space model shown in Eq. \eqref{eq: vechicle}. Here $e_1(t)$, $e_2(t)$, and $\delta(t)$ represent the lateral deviation of the center of mass of the vehicle from the desired path, deviation of yaw angle from the desired yaw angle $\psi_{\rm des}(t)$, and front steering wheel angle, respectively. The desired yaw rate is defined by $\dot{\psi}_{\rm des} = \frac{V_x}{R(t)}$, where $R(t)$ is the radius of a desired path. $C_f$ and $C_r$ are the front and the rear tire cornering stiffness values, respectively. The values of $C_f$ and $C_r$ are uncertain since they are influenced by the tire slip angle. Therefore, instead of assigning them with deterministic values, we treat them as random variables and assume that they satisfy a Gaussian-mixture joint distribution. In this experiment, we assume that they have the same nominal value 967 N/deg. Other parameters of the vehicle model are taken from CarSim (C-Class Hatchback 2017)~\cite{lee2018analysis}. 

\begin{table}[t]
\label{tab:vehicle_model}
 \centering
    \caption{ \fontsize{10}{0}\selectfont Parameters of the vehicle path-following model.}
\begin{tabular}{|c|l|l|}
\hline
Symbol & Description & Value \\ \thickhline
$V_x$ & vehicle speed & 20 m/s\\ \hline
 $m$ &vehicle mass  & 1270 kg\\ \hline
 $a$ & distance from center to front axis & 1.015 m \\ \hline
 $b$ & distance from center to rear axis & 1.895 m \\ \hline
 $I_{zz}$ & vehicle yaw inertia & 1536.7 kg$\cdot$m$^2$ \\ \hline
 $g$ & gravitational acceleration& 9.81 m/s$^2$ \\ \hline

\end{tabular}
\end{table}

\begin{figure*}[t]
    \centering
      \subfigure{
    \includegraphics[width = 3in]{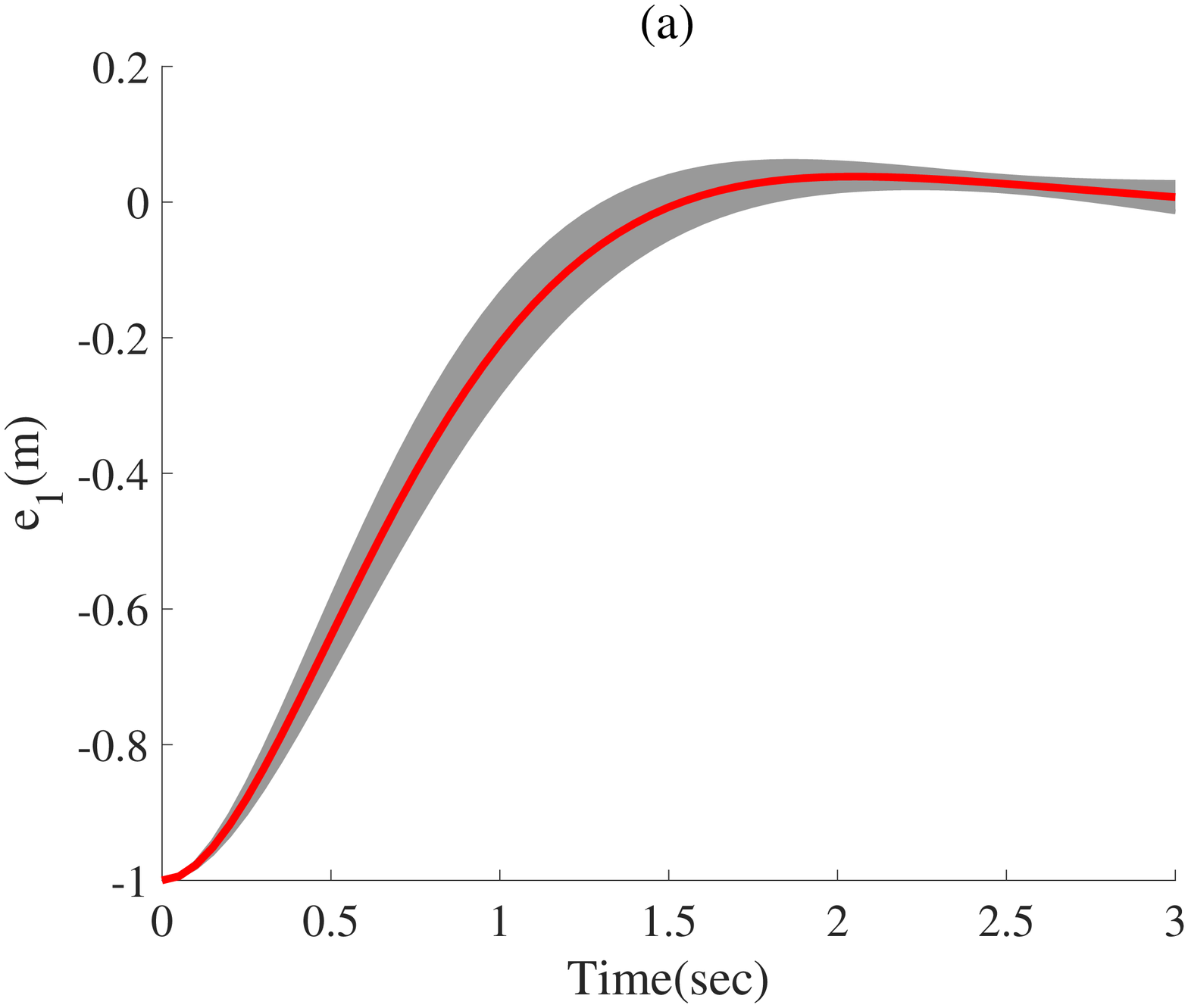}
        
        \label{fig:vehicletraj}
        }
\hspace{0.5pt}
  \subfigure{
    \includegraphics[width = 3in]{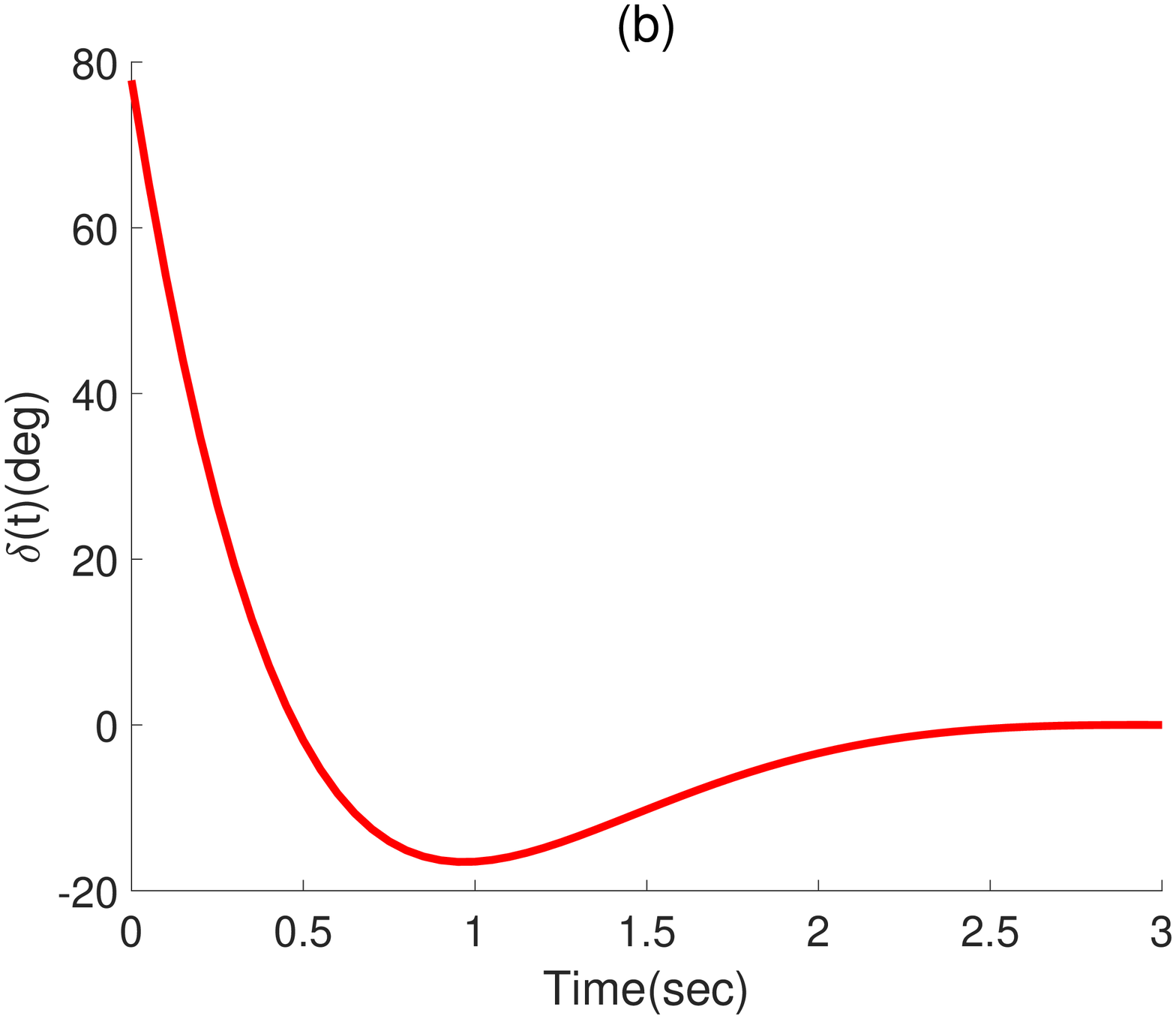}
       
        \label{fig:vehicleinput}
        }
   
    \caption{Numerical results for vehicle path following experiment. (a): resulting lateral position error, where the red line shows the mean value, and the grey lines show the trajectory uncertainties; (b): Optimal control input $\delta(t)$. }
    \label{fig:vehcile}
\end{figure*}

\begin{figure*}[t]
\centering
  \subfigure{\includegraphics[width=3.0in]{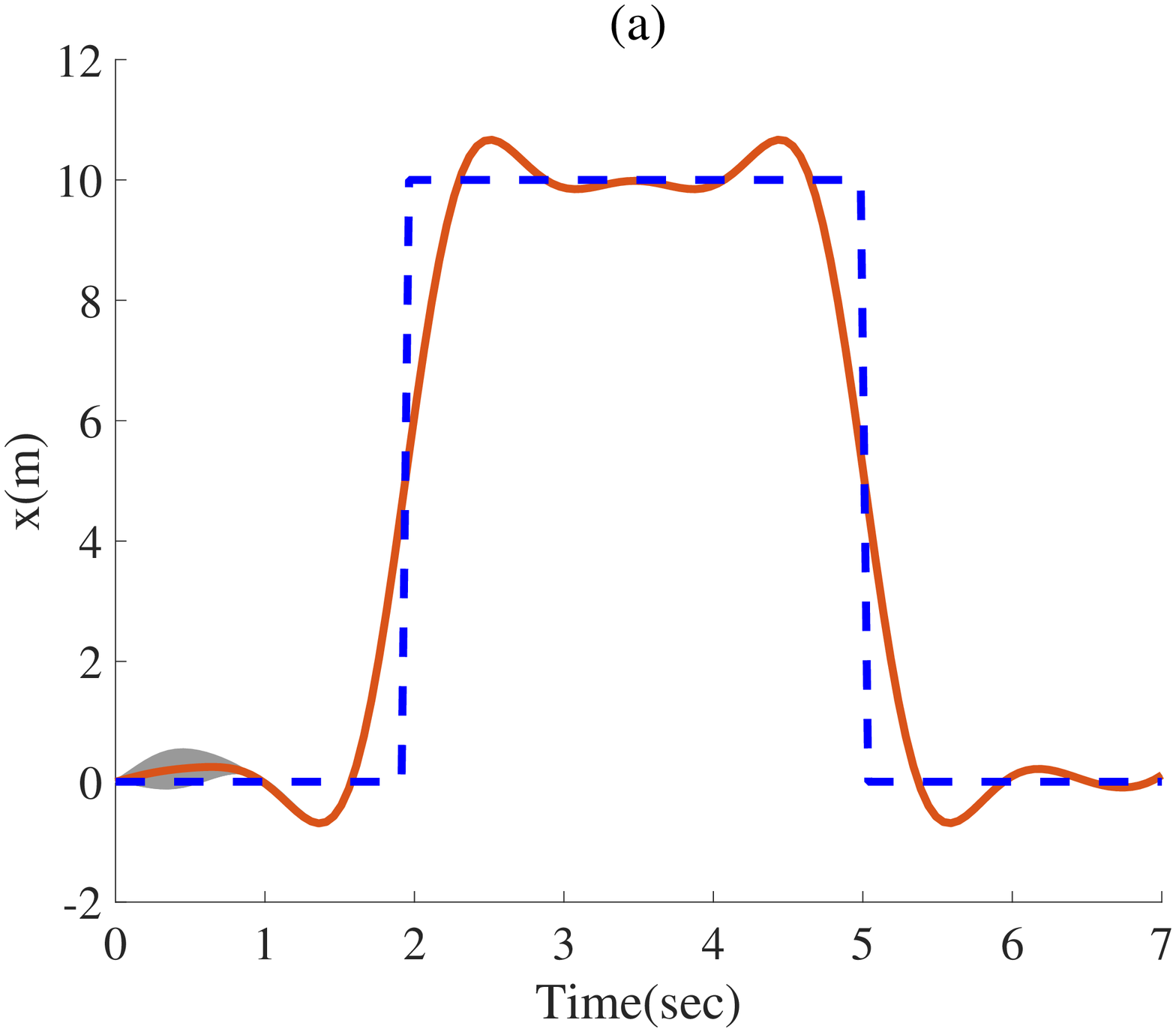}
  \label{fig:stepX}
  }
\hspace{0.5pt}
  \subfigure{\includegraphics[width=3.0in]{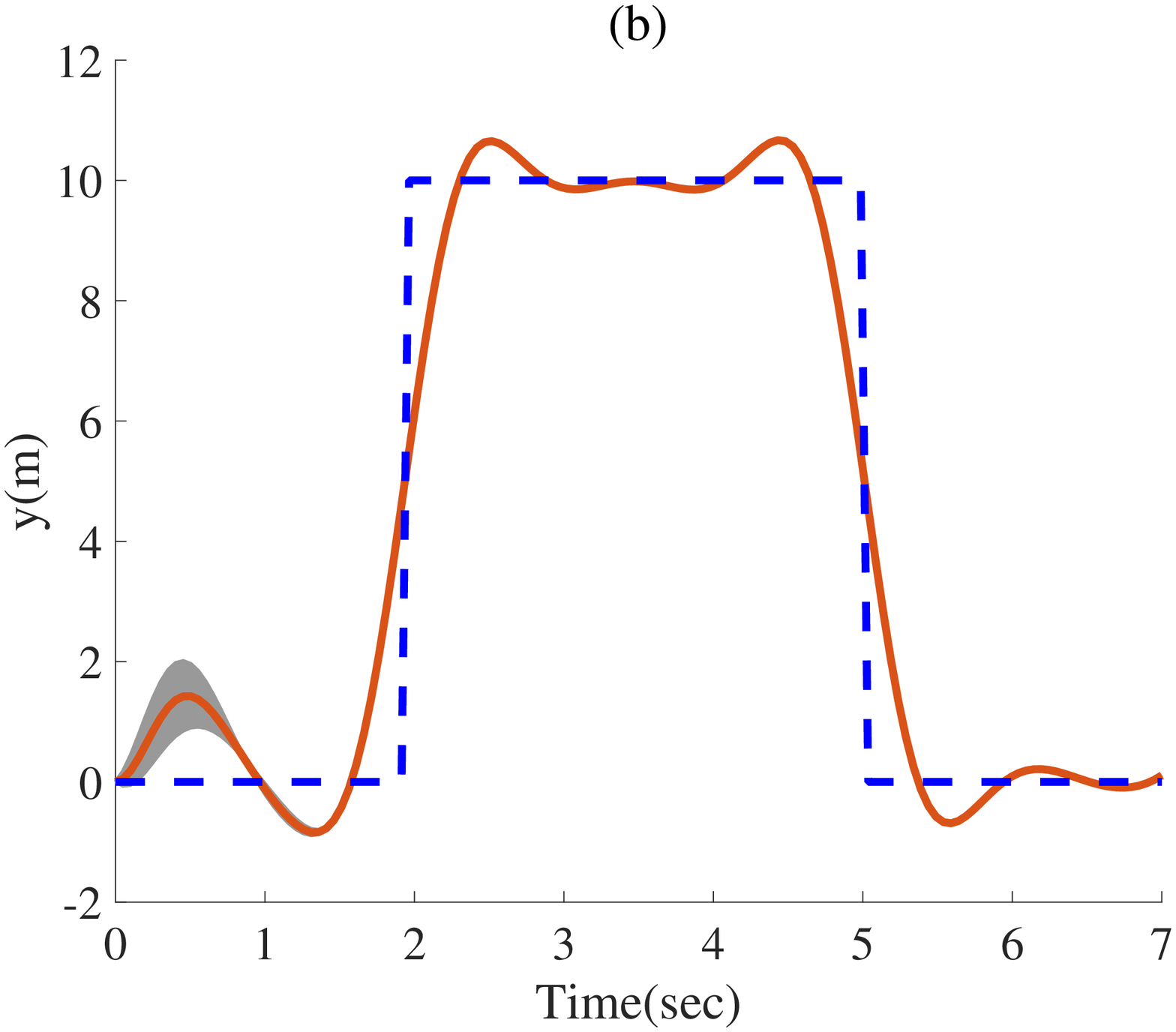}
  \label{fig:stepY}
  }
    \caption{The Quadrotor tracks step reference. (a): in $\mathnormal{x}$ direction; (b): in $\mathnormal{y}$ direction. Red line shows the mean controlled trajectory, the grey line show the effects caused by the uncertainties, and blue dotted line is the reference trajectory.}
    \label{fig:step}
\end{figure*}

Our goal is to enforce the output $\mathbf{y}(t)=e_1(t)$ (i.e., lateral position error) to zero by controlling the steering wheel angle $\delta(t)$. The constraints imposed on the state vector are based on a 99\% level of confidence and are specified below:
\begin{equation}
\begin{aligned} 
 e_1(t) \in [-1,1], \;\; & \dot{e}_1(t) \in [-10, 10], \\
  e_2(t) \in [-28.65, 28.65], \;\; &  \dot{e}_2(t) \in [-572.96, 572.96]. \nonumber
\end{aligned}
\end{equation}
Assuming that the vehicle starts with 1m of a lateral position error, we intend to calculate the optimal steering angle by minimizing the following objective function:
\begin{equation}
    \sum_{t=1}^T \mathbf{x}_t^T\mathbf{Q} \mathbf{x}_t + \mathbf{u}_{t-1}^T\mathbf{R} \mathbf{u}_{t-1} + (\mathbf{y}_t-\mathbf{y}_{{\rm ref},k})^T\mathbf{S}(\mathbf{y}_t-\mathbf{y}_{{\rm ref},k}). \nonumber
\end{equation}
The weighting matrices $\mathbf{Q}$, $\mathbf{R}$, and $\mathbf{S}$ are selected as follows:
\begin{equation}
    \mathbf{Q} = \begin{bmatrix}
    7000 & 0&0&0\\
    0&1&0&0\\
    0&0&20000&0\\
    0&0&0&1
    \end{bmatrix}, \ \ \mathbf{R} = \mathbf{I}, \ \ \ \mathbf{S} = 100. \nonumber
\end{equation}
\par
We employ Alg.~\ref{alg:the_alg} to solve this path-following problem, and the resulting lateral error and control input are plotted in Fig. \ref{fig:vehicletraj} and Fig.~\ref{fig:vehicleinput}, respectively. The lateral position error goes to zeros, which indicates that the vehicle is travelling along the desired path. 


\subsection{Reference Tracking of Quadrotor}
\par
Finally we consider the control of a quadrotor mini-helicopter described in~\cite{lopes2011model}, which consists of four propulsion rotors in a cross configuration. 
The dynamics of this quadrotor can be described by a nonlinear model with 12 state variables:
\begin{equation}
    \dot{\mat{q}} = \mathnormal{f}\left(\mat{q},\mathbf{u} \right),
\end{equation}
where $\mat{q} =\begin{bmatrix} \mathnormal{x},\dot{\mathnormal{x}},\mathnormal{y},\dot{\mathnormal{y}}, \mathnormal{z},\dot{\mathnormal{z}},\phi, \dot{\phi}, \theta, \dot{\theta}, \psi, \dot{\psi}   \end{bmatrix}^T$ is the state vector consisting of Cartesian positions $\mathnormal{x}$, $\mathnormal{y}$, and $\mathnormal{z}$ (in meters), the attitude angles $\phi$ (pitch), $\theta$ (roll), $\psi$ (yaw) in radians, and their respective rates $\left( \dot{\mathnormal{x}}, \dot{\mathnormal{y}}, \dot{\mathnormal{z}}, \dot{\phi}, \dot{\theta}, \dot{\psi} \right)$. Vector $\mathbf{u} =\begin{bmatrix}  \Omega_1,\Omega_2,\Omega_3, \Omega_4 \end{bmatrix}$  is the control input which includes the four propellers' rotational speeds (in radians per second) and $\Omega_i$ is the rotation imposed on the $i$-th motor.
\par
In this experiment, we approximate the quadrotor's dynamics by a linear model around an equilibrium point. The equilibrium point  $\mat{q}_{\rm eq} = \begin{bmatrix} 0,0,0,0,10,0,0,0,0,0,0,0   \end{bmatrix}^T$ and $\mathbf{u}_{\rm eq} = \begin{bmatrix} 192.8, 192.8, 192.8, 192.8  \end{bmatrix}^T$ are used to linearize the system, and this is corresponding to a hover flight condition at a 10m height~\cite{lopes2011model}. With this equilibrium point, the resulting linear state-space model is:

    \begin{equation}
        \begin{aligned}
        &\mat{q}_{t+1} = \mathbf{A}\mat{q}_t+\mathbf{B}\mathbf{u}_t, \; 
        \mathbf{y}_t = \mathbf{C} \mat{q}_t,\\
        \end{aligned}
    \end{equation}
where the model matrices are specified in~\cite{lopes2011model}.   
    

\begin{figure*}[t]
    \centering
      \subfigure{    \includegraphics[width = 3.0in]{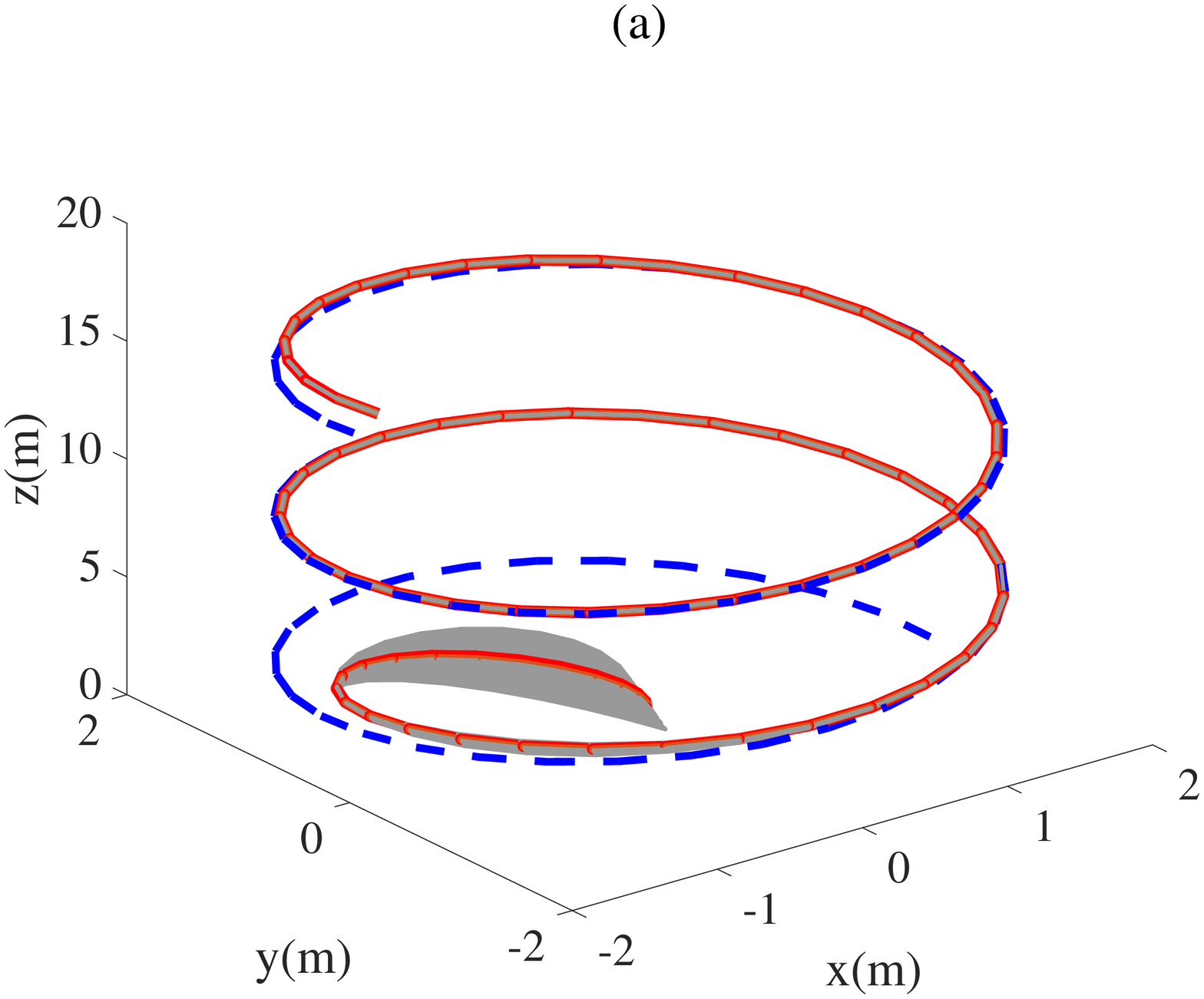}

  \label{fig:defined}
  }
\hspace{0.5pt}
  \subfigure{\includegraphics[width=3.0in]{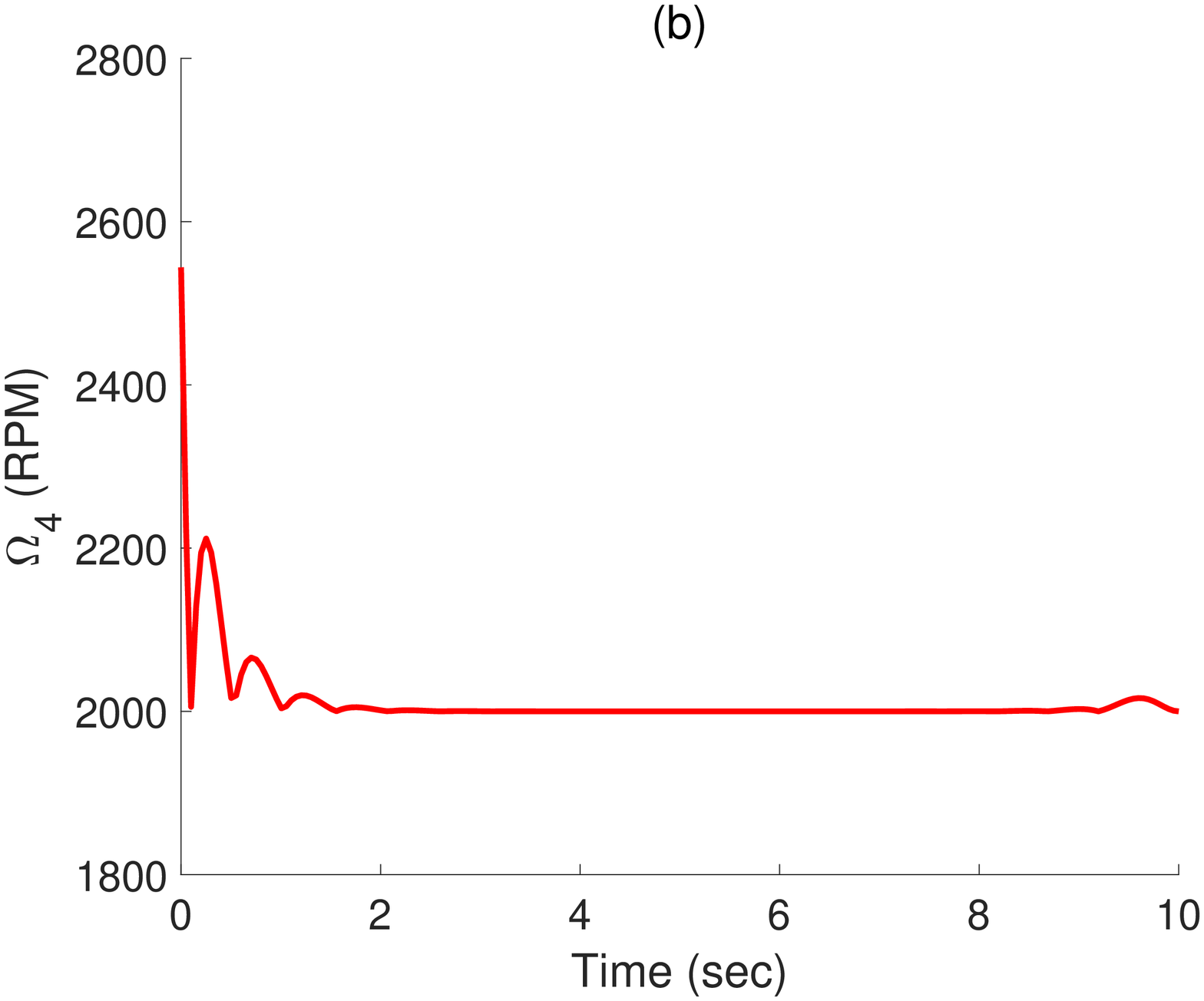}
  \label{fig:quadDefinedInput}
  }
    
    \caption{The quadrotor tracks the reference trajectory in \eqref{eq:ref_traj}. (a): quadrotor trajectories. Red line: mean controlled trajectory. Grey: uncertainties of the controlled trajectory. Blue dotted line: reference trajectory. (b): optimal control input $\Omega_4$.}
    \label{fig:defined}
\end{figure*}

Since there are disturbances acting on the quadrotor, we consider the linear quadrotor system with additive disturbances impose on the three Cartesian positions:
\begin{equation}
  \mat{q}_{t+1} = \mathbf{A} \mat{q}_t+\mathbf{B}\mathbf{u}_t + \mathbf{D}\boldsymbol{\omega}_t(\vecpar),  
\end{equation}
where $\vecpar$ obeys a Gaussian-mixture distribution and matrix $\mathbf{D} \in \mathbb{R}^{12 \times 3}$. 
\par 
The constraints were imposed on $\phi$ and $\theta$ such that $\phi_{\rm max} = 5^{\circ}$, $\phi_{\rm min} = -5^{\circ}$, $\theta_{\rm max} = 5^{\circ}$, and $\theta_{\rm min} = -5^{\circ}$. We first let the quadrotor to track $10$-m step references in the $\mathnormal{x}$ and $\mathnormal{y}$ directions by keeping the elevation at a fixed operating level ($\mathnormal{z} = 10$ m). By using Algorithm \ref{alg:the_alg}, we obtain the controlled $\mathnormal{x}$ and $\mathnormal{y}$ in Fig. \ref{fig:stepX} and \ref{fig:stepY}, respectively. 


\par
Next, we illustrate that the stability of the control loop by letting the quadrotor to track the following reference trajectory:
\begin{equation}
\label{eq:ref_traj}
    \mathnormal{x}_r(t) = 2\cos(0.2t), \ \ \mathnormal{y}_r(t) = 2\sin(0.2t), \ \ \mathnormal{z}_r(t) = 0.2t,
\end{equation}
where $\mathnormal{x}_r$, $\mathnormal{y}_r$, and $\mathnormal{z}_r$ denote the reference trajectories in $\mathnormal{x}$, $\mathnormal{y}$, ans $\mathnormal{z}$ directions, respectively. The quadrotor's behavior following the defined trajectory in $\mathnormal{x}\mathnormal{y}\mathnormal{z}$ plane is shown in Fig. \ref{fig:defined}. The simulation result shows that the controller is able to make the quadrotor follow the desired path successfully.


\subsection{Comparison with Monte Carlo-based MPC}

We compare our proposed algorithm~\ref{alg:the_alg} with Monte Carlo-based MPC approach in terms of CPU time. We use 5000 sample points for Monte Carlo-based MPC. The CPU time required by two methods are summarized in Table~\ref{tab:cpuTime}. Due to the choice of specialized stochastic Galerkin formulation, our method can accurately capture the uncertainties caused by non-Gaussian correlated uncertainties. For instance, Fig.~\ref{fig:obstacleavoidance} shows the PDF of $x_1$ at $t=2$ s for the obstacle avoidance example, where the result from our method is indistinguishable from that of Monte Carlo.

\begin{table}[t]
 \centering
    \caption{\fontsize{10}{0}\selectfont CPU Time comparison of stochastic MPC solvers.}
\begin{tabular}{|l|l|l|}
\hline
   Examples               & Proposed & Monte Carlo \\ \thickhline
Obstacle Avoidance & $250$s & $2.9\times 10^3$s\\ \hline
Vehicle Path-following & $620$s  &  $8.5\times 10^3$s\\ \hline
Quadrotor Tracking A Step Reference& $1300$s & $10^4$s \\ \hline
Quadrotor Tracking Reference \eqref{eq:ref_traj}&  $1200$s & $7000$ s \\ \hline
\end{tabular}
\label{tab:cpuTime}
\end{table}

\begin{figure}[t]
\centering
\includegraphics[width=3in]{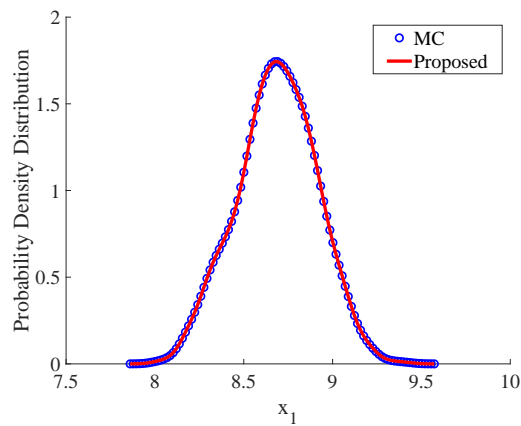}
    \caption{Probability density functions of $x_1$ at $t = 2$ s for the obstacle avoidance example. }
    \label{fig:obstacleavoidance}
\end{figure}

Our method is not compared with existing polynomial chaos-based MPC because the latter cannot handle non-Gaussian correlated uncertainties.

\section{CONCLUSION}
\label{conclusion}
This paper has presented a method for solving chance constrained stochastic MPC problem under non-Gaussian correlated uncertainties. With the proposed stochastic Galerkin formulation, the propagation of uncertainties through the system model can be efficiently and accurately represented. Our framework has reformulated the stochastic system model, objective function and constraints into deterministic ones, resulting in a deterministic optimization problem. This method has been verified on three benchmarks. On these benchmarks, our technique have efficiently handled the non-Gaussian correlated uncertainties and led to a significant reduction of CPU time compared with the Monte Carlo-based model predictive control.   

\section*{Acknowledgement}
The authors would like to thank Dr. Chunfeng Cui and Mr. Zichang He for their technical discussions and for their help in proofreading this manuscript.





\bibliographystyle{IEEEtran}  
\bibliography{Bib/reference} 

\end{document}